\documentstyle[12pt]{article}
\def\beq{\begin{equation}}
\def\eeq{\end{equation}}
\def\bea{\begin{eqnarray}}
\def\eea{\end{eqnarray}}
\def\nn{\nonumber}
\def\ba{\begin{array}}
\def\ea{\end{array}}
\def\d{\partial}
\def\v{\vert}

\def\r{\rangle}
\def\one{1\hskip -1mm{\rm l}}

\begin{document} 
 \begin{flushright} hep-th/9602107
 \end{flushright}
 \rightline{TIFR/TH/96-08}
 \rightline{February 1996} 
\baselineskip16pt
\begin{center} 
{\large \bf \sf
  Spin dependent extension of Calogero-Sutherland model \\ 
  through anyon like representations of permutation operators }\\ 

\vspace{1.5cm}

{\sf B. Basu-Mallick},
\footnote{E-mail address~: biru@theory.tifr.res.in} 

\bigskip

{\em Theoretical Physics Group, \\ 
Tata Institute of Fundamental Research, \\
Homi Bhabha Road, Mumbai-400 005, India} \\

\bigskip

\end{center} 
\vspace {1.75 cm} 
\baselineskip=23pt 
\noindent {\bf Abstract }

We consider a $A_{N-1}$ type of spin dependent Calogero-Sutherland model,
containing an arbitrary representation of the permutation operators on the 
combined internal space of all particles, and find that such a model can 
be solved as easily as its standard $su(M)$ invariant counterpart through 
the diagonalisation of Dunkl operators. A class of novel representations 
of the permutation operator $P_{ij}$, which pick up nontrivial phase factors
along with interchanging  the spins of $i$-th and $j$-th particles, are 
subsequently constructed. These `anyon like' representations interestingly
lead to different variants of spin Calogero-Sutherland model with highly
nonlocal interactions.  We also explicitly derive some exact eigenfunctions 
as well as energy eigenvalues of these models and observe that the related 
degeneracy factors crucially depend on the choice of a few discrete
parameters which characterise such anyon like representations.

\vspace {.75 cm} 

\vspace {.05 cm} 

\newpage 
 
\noindent \section {Introduction }
\renewcommand{\theequation}{1.{\arabic{equation}}}
\setcounter{equation}{0}

\medskip

As it is well known,
the Calogero-Sutherland (CS) model [1,2] and its spin dependent 
generalisations [3-8] fall into a very interesting class of 
quantum many-body systems with long ranged interactions, for  which  
the complete excitation spectrum and various 
dynamical correlation functions can be calculated exactly. Moreover,
in recent years,  such integrable systems  have found a lot of applications 
in apparently diverse subjects like fractional statistics
 in $(1+1)$-dimension [9-13], quantum Hall effect [14-16],
 the level statistics for disordered systems [17-19], matrix models [20,21],
 $W_\infty$ algebra [22-24], etc.

The dynamics of $A_{N-1}$ type spin  CS model,   associated with
 $N$ number of particles   each having $M$ internal degrees of freedom 
 and moving on a ring  of length $L$,  is governed by the
$su(M)$ invariant  Hamiltonian [7,8]
\beq
H ~=~  -{1\over 2} \sum_{i=1}^N ~ \left ( { \d \over \d x_i }
\right )^2 ~+
{ \pi^2 \over L^2 }~ \sum_{i<j}~{   \beta ( \beta + P_{ij}  )  \over
\sin^2 { \pi \over L} (x_i -x_j)  }~.
\label {a1} 
\eeq
Here $\beta $ is a  coupling constant and
$P_{ij}$ is the permutation operator which interchanges the `spins' 
of $i$-th and $j$-th particles. Thus, if the vector $\v {\vec \alpha }\r
~=~ \v \alpha_1 ~\alpha_2  ~\cdots ~ \alpha_N \r $ represents a
particular spin configuration of $N$ particles, with  
$ \alpha_1, ~\alpha_2, ~\cdots ~,~ \alpha_N  ~ \in [1,M] $, 
then $P_{ij}$  will  act on this vector as
\beq 
 P_{ij} ~\v \alpha_1 ~\cdots ~\alpha_i ~\cdots ~ \alpha_j
~\cdots ~ \alpha_N \r ~=~
 \v \gamma_1 ~\cdots ~\gamma_i ~\cdots ~ \gamma_j
~\cdots ~ \gamma_N \r ~,
\label {a2} 
\eeq
where $~\gamma_i  = \alpha_j , ~ \gamma_j  = \alpha_i ~ $ and
$ \gamma_k = \alpha_k $ when $ k \neq i,j $.
The Hamiltonian of the original spin independent CS model [2]
can be recovered from  (\ref {a1})  through the 
formal substitution $P_{ij} \rightarrow -1 $. However it is worth noting
that, in spite their much more complicated nature, 
the eigenstates of the spin CS model (\ref {a1}) can be obtained
 almost in the same way as its spin independent counterpart by
diagonalising a set of simple differential operators known as Dunkl 
operators [7,8]. So it should be interesting to enquire whether there exist
any other form of `permutation' operator $P_{ij}$, than 
given by eqn.(\ref {a2}), which through substitution in
(\ref {a1}) would generate a new quantum Hamiltonian that  can be solved
again through the diagonalisation of these Dunkl operators. 

With the hope of making some progress to the above mentioned 
direction, in sec.2 of this article we briefly recapitulate  the 
procedure of solving the standard $su(M)$ invariant CS model (\ref {a1}). 
In this context we curiously notice that  the algebra of the permutation 
operators $P_{ij}$, rather than any of their particular representation
like (1.2), plays an essential role in solving the model.  
Therefore, if one takes any other representation of $P_{ij}$
on the total internal space of the whole system and substitutes
it to the expression (\ref {a1}), that would also yield a spin 
CS model which can be solved exactly in the same way as its standard
$su(M)$ invariant counterpart.
Next, in sec.3, we construct a new class of representations
of the  permutation operator $P_{ij}$, 
by considering a specific limit of some known  
braid group representations associated with the universal 
${\cal R}$-matrix of $U_q(sl(M))$ quantum group.
Such  novel representations of the permutation operators,
characterised by a set of discrete as well as continuous deformation 
parameters, interestingly lead to different types 
of exactly solvable spin CS models  with highly nonlocal interactions 
which would break the $su(M)$ invariance.
Subsequently, in sec.4, we focus our attention to 
some special cases of these  models
containing a small number of spin-${1\over 2}$ particles 
and explicitly derive the related  eigenvectors for several
 low-lying energy states. Sec.5 is the concluding section.

\vspace{.75cm}

\noindent \section {Solution of $su(M)$ invariant CS model }
\renewcommand{\theequation}{2.{\arabic{equation}}}
\setcounter{equation}{0}

\medskip

To  solve the $su(M)$ invariant CS model (\ref {a1}), 
it is useful to make an ansatz
for  the corresponding wave function as [8]
\beq
 \psi ( x_1 , ~ \cdots ~ ,~ x_N ;~\alpha_1,~\cdots ~,~
 \alpha_N ) ~~=~~ \Gamma^\beta ~ 
 \phi ( x_1 , ~ \cdots ~ ,~ x_N ;~\alpha_1,~\cdots ~,~
 \alpha_N )~,
\label{b1}
\eeq
where $\Gamma = \prod_{i<j} \sin { \pi \over L} (x_i -x_j ) $
(here $\beta $ is assumed to be positive
 for avoiding singularity at $ x_i \rightarrow x_j $). Next, by applying
the canonical commutation relations 
$ [ { \d\over \d x_j } , x_k ] = \delta_{jk} $
 and also making a change of
coordinates like $z_j = e^{{ 2\pi i \over L } x_j } $, one may easily find
that
\beq
  H \psi ~=~ H \Gamma^\beta \phi ~=~
 {2\pi^2 \over L^2 }~ \Gamma^\beta ~{\cal H} \phi ~,
\label{b2}
\eeq
where $H$ is the original Hamiltonian (\ref {a1}) and 
\bea 
 &  & {\cal H}  ~=~ \sum_j \left ( z_j { \d \over \d z_j } \right )^2 ~+~ 
\beta ~\sum_{ i<j} ~ { z_i + z_j \over z_i - z_j } ~
\left ( z_i { \d \over \d z_i } - z_j { \d \over \d z_j } \right ) \nn \\
  &   & \hskip 3.5 cm
- ~2 \beta  ~ \sum_{i <j} ( 1 + P_{ij} )
 { z_i  z_j \over ( z_i - z_j)^2 } ~+~ { \beta^2 \over 12 } N ( N^2 -1 )~.
\label {b3} 
\eea 
Due to the `gauge transformation' (\ref {b2}), 
the diagonalisation problem of $H$
is now reduced to the diagonalisation problem of effective
Hamiltonian  $ {\cal H}$.
Thus, if $\phi $ is an eigenvector of ${\cal H}$ with eigenvalue $\epsilon$,
 then $\psi $ would be the corresponding
eigenvector of $H$  satisfying the relation
\beq
  H \psi ~=~ { 2 \pi^2 \over L^2 } \epsilon ~ \psi ~. 
\label {b4} 
\eeq  
To diagonalise $ {\cal H} $, however, it is convenient to introduce
another operator ${\cal H}^*$ which acts only on the coordinate degrees
of freedom and may be given by
\bea 
 &  &  {\cal H}^* ~=~ \sum_j \left ( z_j { \d \over \d z_j } \right )^2 ~+~ 
\beta ~\sum_{ i<j} ~ { z_i + z_j \over z_i - z_j } ~
\left ( z_i { \d \over \d z_i } - z_j { \d \over \d z_j } \right ) \nn \\
  &   & \hskip 3.5 cm
- ~2 \beta  ~ \sum_{i <j} ( 1 - K_{ij} )
 { z_i  z_j \over ( z_i - z_j)^2 } ~+~ { \beta^2 \over 12 } N ( N^2 -1 )~,
\label {b5} 
\eea 
where $K_{ij}$s are the coordinate exchange operators defined through
algebraic relations
\bea
& \quad \quad  K_{ij} z_i ~=~ z_j K_{ij},~~K_{ij} { \d \over \d z_i } ~=~ 
{ \d \over \d z_j } K_{ij}, ~~K_{ij} z_l ~=~ z_l K_{ij}~,~ \nn &
\quad \quad  \quad   (2.6a)  \\ 
& \quad \quad  K_{ij}^2 ~=~ 1, ~~~K_{ij}K_{jl} ~=~ K_{il}K_{ij} ~=~
K_{jl} K_{il} ~, ~~~[\, K_{ij}, K_{lm} \, ] ~=~ 0~,~ \nn  &  
\quad  \quad \quad   (2.6b)
\eea 
\addtocounter{equation}{1} 
\noindent $ i, ~j,~l,~m ~$ being all different indices. 
It may be noted that  the  operator 
${\cal H} $ in eqn.(\ref {b3}) can be reproduced  from the expression of 
 ${\cal H}^*$ in (\ref {b5}) 
through the formal substitution $K_{ij} \rightarrow 
- P_{ij} $. Due to such close connection between these two operators
and also because ${\cal H}^*$ satisfies some simple commutation relations like
\beq
  \left [\, {\cal H}^*  ,~ K_{ij} \, \right  ]  
~=~ \left [ \, {\cal H}^* ,~ P_{ij} \, \right  ] ~=~ 0 ~,
\label {b7} 
\eeq
one can easily
 construct an eigenfunction  of $ {\cal H} $ from a given
eigenfunction  of ${\cal H}^*$ in the following way.  

Let $ \Pi_{ij} $s be
the set of permutation operators  which simultaneously interchange
the spins as well as coordinates of two particles (i.e., 
$ \Pi_{ij} = P_{ij} K_{ij} $) and 
$ \Lambda $ be the corresponding 
antisymmetric projection operator satisfying the relations
\beq 
 \Pi_{ij}  \Lambda ~=~ P_{ij} K_{ij}  \Lambda  ~=~ - ~\Lambda ~,
\label {b8} 
\eeq
(or, equivalently, $ P_{ij } \Lambda = - K_{ij} \Lambda $). 
Now, if $ \xi \equiv \xi 
( z_1, z_2, \cdots , z_N )$  is  an eigenvector 
of ${\cal H}^*$ with eigenvalue $\epsilon $, then by using commutation
relations  (\ref {b7}) 
one can generate  another eigenvector of ${\cal H}^*$  with 
the same eigenvalue:
$ {\cal H}^* \Lambda ( \xi \rho ) ~=~ \epsilon ~\Lambda ( \xi \rho ) ~,
$ where $\rho \equiv \rho ( \alpha_1, \alpha_2, \cdots ,\alpha_N ) $
is an arbitrary spin dependent function. However, due to property
(\ref {b8}) of the  antisymmetriser $\Lambda $, it is evident that
$ \Lambda  ( \xi \rho )$ may also be considered as an eigenfunction of
the effective Hamiltonian $ {\cal H} $ with eigenvalue $\epsilon $. 
As a result the eigenvectors of original
Hamiltonian  $H$, defined through  eqn.(\ref {b4}), 
can be written  through the eigenfunctions of ${\cal H}^*$ as
\beq 
\psi ~~=~~ \Gamma^\beta ~ \Lambda ( \xi \rho ) ~.
\label {b9} 
\eeq 

Finally,  for finding out the eigenfunctions of ${\cal H}^*$,  
it may be observed  that  this operator  can be expressed in
a simple quadratic  form like [8]
\beq
{\cal H}^* ~~=~~ \sum_{i=1}^N ~ d_i^2 ~,
\label {b10}
\eeq
where $d_i$s $( i \in [1,N] )$ are the so called Dunkl operators [25-27]:
\beq
d_i ~=~ z_i { \d \over \d z_i } ~+~ 
\beta \left (~ i - { N+1 \over 2 } ~\right )
~-~ \beta ~\sum_{j>i } { z_i \over z_i - z_j } \left ( K_{ij} -1 \right )
~+~ \beta ~\sum_{j<i } { z_j \over z_j - z_i }  \left ( K_{ij} -1 \right ) ~,
\label {b11}
\eeq
which satisfy the relations
\beq 
\left [ ~d_i \,  , \,  d_j ~\right ] ~=~0 ~,~~
\left [ ~k_{i, i+1 } \,  , \,  d_l ~\right ] ~=~0 ~,~~
k_{i, i+1 }  d_i ~-~ d_{i+1}  k_{i,i+1 }  ~=~ -  \beta  ~, 
\label {b12}
\eeq
when  $ l \neq i,~i+1 $. So  one should be able to
 construct the eigenvectors
of ${\cal H}^*$ by simultaneously diagonalising the  mutually commuting 
operators $d_i$.  For this purpose, however,
it is quite  helpful to make the following 
ordering  of the corresponding  basis
elements  characterised by the monomials 
\beq 
m_{ \left \{
 \lambda_1 , \lambda_2 , \cdots , \lambda_N  \right \} }~=~ z_1^{\lambda_1 }
z_2^{\lambda_2 } \cdots z_N^{\lambda } ~,
\label {b13}
\eeq
where $\{ \lambda_1 , \cdots , \lambda_N \} \equiv  [\lambda ] $ 
is a sequence of non-negative integers
with the homogeneity  $ \lambda = \sum_{i=1}^N \lambda_i $. 
 For this sequence
$[\lambda]$, one may  now associate a partition $\v \lambda \v $ 
where the  entries $ \lambda_k $ are arranged in decreasing order.
Next,  an ordering among the partitions (which are obtained
from all monomials with a given homogeneity $\lambda $) is defined
by saying that $\v \lambda \v $ is larger than $ \v \mu  \v$  if 
the first nonvanishing difference $\lambda_k - \mu_k $ is positive.
This prescription naturally induces an ordering between any two
monomials which belong to different partitions
of homogeneity $\lambda $. One can further
order the monomials associated with the same  partition by saying that 
$[\lambda ] $ is larger than $ [ \lambda' ] $ if the last non-vanishing
difference $\lambda_k - \lambda'_k $ is positive. Due to such  global
ordering it turns out that the  operators $d_i$ and ${\cal H}^*$ can be written
through block-triangular matrices; each block representing the action 
of an operator  on all  monomials
within a given  homogeneity sector [7]. 
By using this  important block-triangular property, for which the 
diagonal elements of a matrix can be identified with the corresponding 
eigenvalues, one finds  that
\beq
{\cal H}^* ~\xi_{ \left \{
 \lambda_1 , \lambda_2 , \cdots , \lambda_N \right \} } ~=~ 
\sum_{i=1}^N ~
 \left [ \lambda_i - \beta \left ( { N+1 \over 2 } -i \right ) 
\right ]^2 ~
 \xi_{ \left \{  \lambda_1 , \lambda_2 , \cdots , \lambda_N  \right \} }
\label {b14}
\eeq
where the eigenfunction 
 $ \xi_{ \left \{ \lambda_1 , \lambda_2 , \cdots , \lambda_N \right  \} } $ 
is a suitable linear combination of 
$ m_{ \left  \{ \lambda_1 , \lambda_2 , \cdots , \lambda_N  \right  \} }$ 
and other monomials of relatively lower orders. 

Though it is rather difficult to 
write down the general form of
$\xi_{\left \{ \lambda_1 , \lambda_2 , \cdots , \lambda_N \right \} } $,
one can find it out easily  for the  case of
low-lying energy states
associated with small number of particles. We  shall explicitly derive  a
few of such $ \xi_{ \left \{ 
\lambda_1 , \lambda_2 , \cdots , \lambda_N  \right \} } $
in sec.4 of this article,  and subsequently use them to  generate
the eigenstates  of  new spin  CS models  which are related to
some  `anyon like' representations of the  permutation operators.

\vspace{.75cm}

\noindent \section { Novel variants of spin CS model } 
\renewcommand{\theequation}{3.{\arabic{equation}}}
\setcounter{equation}{0}

\medskip
In close analogy with the 
$su(M)$ invariant CS model (\ref {a1}), 
we consider in the following another Hamiltonian ${\tilde H}$  which can
differ from (\ref {a1}) only  through the nature of corresponding 
spin-spin interactions.
\beq
{\tilde H} ~=~  -{1\over 2} \sum_{i=1}^N ~\left (
 { \d \over \d x_i } \right )^2 ~+
{ \pi^2 \over L^2 }~ \sum_{i<j}~
{   \beta ( \beta + {\tilde P}_{ij}  )  \over
\sin^2 { \pi \over L} (x_i -x_j)  }~,
\label {c1} 
\eeq
where ${\tilde P}_{ij}$s  are  any possible set of
`permutation'  operators which  act on the combined
internal space of $N$ particles (i.e., on $ {\cal F} ~\equiv ~
 \underbrace { 
C^M \otimes C^M \otimes  \cdots \otimes C^M }_N ~$) 
and yield a representation of the algebraic relations 
\beq 
  {\cal P}_{ij}^2 ~=~ 1, 
~~~{\cal P}_{ij}{\cal P}_{jl} ~=~ {\cal P}_{il}{\cal P}_{ij} ~=~
{\cal P}_{jl} {\cal P}_{il} ~, ~~~[ {\cal P}_{ij}, 
{\cal P}_{lm} ] ~=~ 0~ ,
\label {c2}
\eeq 
$ i, ~j,~l,~m $  being all different indices. 

It should be noticed
that the standard permutation operator $P_{ij}$, defined by
eqn.(\ref {a2}), is only a particular representation  of 
${\cal P}_{ij}$  satisfying the algebra (\ref {c2}). Our aim is to
construct here some other 
 representations of ${\cal P}_{ij}$ on the vector space $ {\cal F}$
and subsequently use such ${\tilde P}_{ij}$  to generate new variants 
of spin CS model. However, before
focussing our attention to  those specific cases,  let us 
 investigate at present  how a Hamiltonian like (\ref {c1}), containing
an arbitrary representation of permutation operators,  can
be solved exactly by using the techniques which  
 have been already discussed in sec.2.
For this purpose we assume that the form of 
the corresponding wave function ${\tilde \psi} $ is
 again given by an  ansatz like (\ref {b1}) 
and make the `gauge transformation': 
  $~ {\tilde H} {\tilde \psi } ~=~ {\tilde H}  \Gamma^\beta {\tilde \phi}
~=~ {2\pi^2 \over L^2 }~ \Gamma^\beta ~{\tilde {\cal H }} {\tilde \phi },~$
where the  effective Hamiltonian $ {\tilde {\cal H}} $  is now
expressed as
\bea 
 && {\tilde {\cal H}}  ~=~ \sum_j \left ( z_j { \d \over \d z_j } \right )^2 
~+~ \beta ~\sum_{ i<j} ~ { z_i + z_j \over z_i - z_j } ~
\left ( z_i {  \d \over \d z_i } - z_j { \d \over \d z_j } \right ) \nn \\
  & & \hskip 3.5 cm
- ~2 \beta  ~ \sum_{i<j} ( 1 + {\tilde P}_{ij} )
 { z_i  z_j \over ( z_i - z_j)^2 } ~+~ { \beta^2 \over 12 } N ( N^2 -1 )~.
\label {c3} 
\eea 
Evidently, the above effective Hamiltonian $ {\tilde {\cal H}}$ 
can  also be reproduced from the operator ${\cal H}^*$ in eqn.(\ref {b5}), 
through the formal substitution: $ K_{ij} \rightarrow - {\tilde P}_{ij}$.
So, for constructing the eigenvectors of $ {\tilde {\cal H}}$ from that of
${\cal H}^*$, we define a set of operators as $  {\tilde \Pi}_{ij} ~=~K_{ij} 
{\tilde P}_{ij} $.  Since  both $K_{ij}$ and ${\tilde P}_{ij}$ satisfy 
an algebra like (\ref {c2}), while  acting on the coordinate
and spin spaces respectively, the newly defined operators 
$ {\tilde \Pi}_{ij} $ would also
produce a representation of the same permutation algebra 
on the full Hilbert space of $N$ particles.
Therefore, by using only this permutation algebra,
 one can easily define a `generalised'
antisymmetric  projection operator ${\tilde \Lambda } $ 
which will satisfy the relations
\beq 
 {\tilde \Pi}_{ij} {\tilde \Lambda }
  ~=~ K_{ij} {\tilde P}_{ij} {\tilde  \Lambda }  ~=~ - ~{\tilde \Lambda } ~.
\label {c4} 
\eeq
For example, such antisymmetric projection operators corresponding
to the simplest $N=2 $ and $N=3$ cases
(denoted by ${\tilde \Lambda }_2 $ and ${\tilde \Lambda }_3 $
respectively)  are given by
$$
 {\tilde \Lambda}_2 ~=~ 1 - {\tilde \Pi }_{12} ~,~~
 {\tilde \Lambda}_3 ~=~1 - {\tilde \Pi}_{12} - {\tilde \Pi}_{23}
 - {\tilde \Pi}_{13} + {\tilde \Pi}_{23} {\tilde \Pi}_{12}  +
{\tilde \Pi}_{12} {\tilde \Pi}_{23} ~.  \eqno (3.5a,b) 
$$
\addtocounter{equation}{1} \noindent
Now, by exactly following the arguments of sec.2,  it is straightforward
to verify  that 
$~ {\tilde \Lambda }  ( \xi \rho ) ~$ will  be  an 
  eigenvector of $ {\tilde {\cal H}}$  with eigenvalue $\epsilon $,
when  $ \xi \equiv \xi ( z_1, z_2, \cdots , z_N )$ 
 is an eigenfunction of ${\cal H}^*$ with  same eigenvalue, and 
 $\rho \equiv $ $\rho ( \alpha_1,$ $  \alpha_2,$ $ \cdots ,\alpha_N ) $
is an arbitrary spin dependent function. Finally, 
by using a relation like (\ref {b2}), the eigenfunction of
  spin  CS model (\ref {c1}) corresponding to
eigenvalue ${2\pi^2\over L^2 } \epsilon~ $ can also  be expressed  through 
the eigenfunction  of ${\cal H}^*$  as
\beq
{\tilde \psi } ~~=~~ \Gamma^\beta ~
{\tilde \Lambda } \,  ( \xi \rho ) ~.
\label {c6}
\eeq

Thus from the above discussion it is clear that the spin CS 
  Hamiltonian (\ref {c1}) can be solved much in the same 
way as its original counterpart (\ref {a1}) by introducing a
generalised antisymmetric projection operator ${\tilde \Lambda}$.
The key point in this approach is that for  finding out 
the form of ${\tilde \Lambda}$, which satisfies  equation (\ref {c4})
and consequently  relates  the eigenvectors of 
${\cal H}^*$  and  ${\tilde {\cal H}}$,  
one does not have to use any information about the representation
 ${\tilde P}_{ij}$ except
 that it satisfies the permutation algebra (\ref {c2}). 
So, each representation
of ${\cal P}_{ij}$ on the combined internal space of $N$ particles 
 would generate an exactly solvable model
whose eigenvectors can be obtained by using the equation (\ref {c6}).
The standard representation (\ref {a2}) evidently
reproduces the well known $su(N)$ invariant CS Hamiltonian (\ref {a1}) 
with eigenvectors given by eqn.(\ref {b9}).

Now, for constructing other possible representations of the  
permutation group ($S_N$) related algebra
(\ref {c2}), we  recall that it  can be  generated by 
$N-1$  number of elements
${\cal P}_{k,k+1}$ $(k \in [1,N-1] )$,
which satisfy the relations 
\bea 
&& {\cal P}_{k,k+1} {\cal P}_{k+1,k+2} {\cal P}_{k,k+1} \,  = \,
{\cal P}_{k+1,k+2} {\cal P}_{k,k+1} {\cal P}_{k+1,k+2} \, ,~
\left [  {\cal P}_{k, k+1 },{\cal P}_{l, l+1} \right ] \, = \, 0\, , ~
 {\cal P}_{k,k+1}^2 \, = \,  \one \, ,~~~~~~~~ \nn\\
&& \hskip 12.85  cm \nn  (3.7a,b,c) 
\eea
\addtocounter{equation}{1}
where  $\v k-l \v > 1 $. All  other `non-nearest neighbour'
elements like  ${\cal P}_{km} $ (with $m-k >1 $) can be expressed
through these generators as 
\beq
{\cal P}_{km} ~~=~~ \left ( 
{\cal P}_{k,k+1} {\cal P}_{k+1,k+2} \cdots  {\cal P}_{m-2,m-1} \right )
\, {\cal P}_{m-1,m} \, \left ( {\cal P}_{m-2,m-1} \cdots 
{\cal P}_{k+1,k+2} {\cal P}_{k,k+1} \right ) ~.
\label {c8}
\eeq
It is worth observing 
in this context that the braid group $B_N$ [28] also has 
 $N-1$ number of generators $b_{k}$ ($k \in [1,N-1]$) and
the corresponding algebra  looks  very similar to  the 
relations  (3.7a,b):
\beq 
b_k b_{k+1} b_k ~=~b_{k+1} b_k b_{k+1} ~,~~~
\left [ \,  b_k \, , \, b_l \, \right ] ~=~0~ 
\label {c9}
\eeq
 where  $\v k -l  \v > 1 $.  However,
 there is no analogue of 
the relation (3.7c) for the braid group generators. To `reduce'
 this difference between the generators of $S_N$ and $B_N$,
one may consider a specific  class of braid group representations
(BGRs) which satisfy the extra  condition
\beq 
  b_k^2 ~~=~~ ( q - q^{-1} )~b_k ~+~ \one ~, 
\label {c10}
\eeq
$q$ being an arbitrary nonvanishing parameter. In fact,  the 
equations  (\ref {c9}) and (\ref {c10}) define  together
the Hecke algebra, which has interesting applications 
in many  areas related to integrable models [28-30]. 
For the present purpose
it is useful to notice that at the limit $ q\rightarrow 1 $,
 eqn.(\ref {c10}) becomes exactly equivalent to the relation
(3.7c). Consequently, by taking this limit  to some known representations 
of Hecke algebra and making the identification
 $b_k \rightarrow  {\cal P}_{k,k+1} $, we might be able to construct  new
  representations of algebra (3.7)  satisfied by the permutation
generators.

The representations of braid group (and also of Hecke algebra at some
special cases),  in turn,
can be derived in a systematic  way by using the universal
${\cal R}$-matrix  associated with various  quantum groups [31-33].
A class of such BGRs, operating on the 
tensor product space ${\cal F}$, are given by
\beq
b_k ~~=~~ \sum_{\sigma = 1}^M ~   \epsilon_\sigma (q) ~
e^k_{\sigma \sigma }  \otimes   e^{k+1}_{\sigma \sigma } ~+~
\sum_{ \sigma \neq \gamma } ~e^{ i \phi_{ \gamma  \sigma } }~ 
e^k_{\sigma \gamma } \otimes  e^{k+1}_{\gamma \sigma } ~+~
\left ( q - q^{-1} \right )~ \sum_{  \sigma  < \gamma } ~
e^k_{\gamma \gamma } \otimes  e^{k+1}_{\sigma \sigma } ~,
\label {c11}
\eeq
where $ e^k_{\sigma \gamma } $ are the basis operators on the $k$-th 
vector space with elements
 $ \left ( e^k_{\sigma \gamma } \right )_{ \tau \delta }
= \delta_{\sigma \tau } \delta_{ \gamma \delta } $,  
$ \, \phi_{ \gamma  \sigma }$ are ${ M(M-1) \over 2 }$ number of 
independent  antisymmetric deformation parameters: 
$ \phi_{ \gamma  \sigma }  =  - \phi_{ \sigma  \gamma } $, 
and each of the $  \epsilon_\sigma (q) $ can be  freely 
 taken as either $ q$ or $- q^{-1} $ for any value of $\sigma $. 
So for every  possible choice
of the set of parameters $q, ~\phi_{ \sigma \gamma }$ and 
$  \epsilon_\sigma (q) $,   eqn.(\ref {c11}) will give us
 a distinct braid group representation. 
Though the  derivation of  relation (\ref {c11}) is not relevant
for our purpose, it is worth noting that in the special  case  when 
 all  $\epsilon_\sigma (q)$s  take the same value
(i.e., all of them are either $q$ or $-q^{-1}$), the  corresponding
BGRs can be obtained from the fundamental representation of the universal
${\cal R}$-matrix associated with  $U_q(sl(M))$ quantum group, for generic 
values of the parameter $q$ [31-33]. On the other hand if 
 $\epsilon_\sigma (q)$s  do not take the same value for all $\sigma $, 
the corresponding `nonstandard' BGRs  are found to be connected with
 the universal ${\cal R}$-matrix of $U_q(sl(M))$ quantum group
 when    $q$ is  a root of unity [34-36].  Furthermore, the parameters
$\phi_{\sigma \gamma }$ and $\epsilon_\sigma (q)$ have appeared
 previously in the context of  multi-parameter dependent quantisation of 
$GL(M)$ group [37] and the asymmetric vertex model studied by Perk
and Schultz [38]. However, one may also directly check that
the BGRs given by eqn.(\ref {c11})
obey both the relations (\ref {c9}) and (\ref {c10}), and therefore,
can be considered as some  representations of the Hecke algebra. 
Consequently, by taking the $ q\rightarrow
1 $ limit of the expression (\ref {c11}),   we get a class of
representations of the permutation algebra (3.7) as
\beq
{\tilde P}_{k,k+1} ~=~ \sum_{\sigma = 1}^M ~   \epsilon_\sigma  ~
e^k_{\sigma \sigma }  \otimes   e^{k+1}_{\sigma \sigma } ~+~
\sum_{ \sigma \neq \gamma } ~e^{ i \phi_{ \gamma  \sigma } }~ 
e^k_{\sigma \gamma } \otimes  e^{k+1}_{\gamma \sigma } ~,
\label {c12}
\eeq
where $\epsilon_\sigma  $ can be freely  
chosen to be either $1$ or $-1$ for each value of $\sigma $. 
By inserting the above expression of ${\tilde P}_{k,k+1}$
to eqn.(\ref {c8}), one can easily construct
the representations  of `non-nearest
neighbour' permutation elements.  It might  be noted that, 
the parameters $\epsilon_\sigma $
in eqn.(\ref {c12}) have some apparent  similarity 
with the grading parameters which appear in the supersymmetric
 exchange operator and related CS model [39]. 
However, in contrast to the case of ref.39, our  
expression  (\ref {c12}) defines some representations
of permutation algebra (3.7) on the usual vector space ${\cal F}$
 which does not carry any $Z_2$ grading.

It is evident that when
 $ \, \epsilon_\sigma =1 , \, $$\phi_{\gamma \sigma } = 0 $   
 for all values of  $\sigma ,  \, \gamma $,
the expression (\ref {c12}) coincides with the standard form of 
permutation operator  (\ref {a2}). Similarly,  the case
$  \, \epsilon_\sigma = - 1 , \, $$\phi_{\gamma \sigma } = \pi $ 
 for all  $\sigma , ~\gamma $  reproduces again
 the representation (\ref {a2})
 up to an over all sign factor. However to get an
insight to  other situations, let us consider first the simplest case
of  two spin-${1\over 2}$ particles (i.e,  $N=M=2$), where
we have only one permutation operator ${\tilde P}_{12}$. 
By using the relation (\ref {c12}), one can explicitly  write down the
action of this operator on  the associated vector as
\bea 
{\tilde P}_{12} \v 11 \r \, = \,  \epsilon_1 \v 11 \r \,  ,~
{\tilde P}_{12} \v 12 \r \,   =  \,  e^{i \theta } \v 21 \r \,  ,~ 
{\tilde P}_{12} \v 21 \r \,  =  \,   e^{-i \theta } \v 12 \r \,  ,~
{\tilde P}_{12} \v  22 \r \,   = \,  \epsilon_2 \v 22  \r \,  ,~
\label {c13}
\eea
where $\theta = \phi_{12} $. It is curious to notice that,
somewhat similar to the case of anyons,
the above representation of ${\tilde P}_{12}$  not only 
interchanges the spin of two particles but also picks 
up some spin-dependent phase factors. Consequently,
when substituted in the Hamiltonian (\ref {c1}),  such representation 
of permutation operator would  break the $su(2)$ 
symmetry which is present in the original  spin-${1\over 2}$ CS model.

More interesting things 
happen if one considers the motion of three spin-${1\over 2}$ particles 
(i.e. $N=3, \, M=2$). In this case we naturally have three permutation 
operators ${\tilde P}_{12},$ ${\tilde P}_{23}$ and  ${\tilde P}_{13}$, 
which act on the direct product of three spin spaces.
However, the forms of ${\tilde P}_{12}$ and  ${\tilde P}_{23}$ 
would again be given by  equations like (\ref {c13}) while these operators
act on the direct product of two spin spaces where they are nontrivial. 
To find out the action of  the remaining element
${\tilde P}_{13}$, we have to simultaneously use
the above mentioned  forms  of 
${\tilde  P}_{12}, ~ {\tilde  P}_{23}$ and the  relation
 ${\cal P}_{13} \,  = \, {\cal P}_{12} {\cal P}_{23}
 {\cal P}_{12}$ (which is derivable from (\ref {c8})):
\bea
\quad  \quad \quad  &{\tilde P}_{13} \v \, 1 \alpha_2 1 \, \r  ~=~
 \epsilon_1 \v \,  1 \alpha_2 1 \,  \r ~ ,~~~
{\tilde P}_{13} \v  \, 1 \alpha_2  2 \,  \r ~ = ~  f(\alpha_2) \, 
\v \,  2 \alpha_2 1 \, \r ~ , \nn   \hskip 1.30 cm  &(3.14a,b) \\
  \quad  \quad \quad  &{\tilde P}_{13} \,  \v  \,
 2 \alpha_2 1 \, \r ~=~  g(\alpha_2) \,
\v \,  1 \alpha_2 2 \,  \r ~ ,  ~~~
{\tilde P}_{13} \,  \v \, 2 \alpha_2 2 \,  \r ~ = ~
 \epsilon_2  \, \v  \, 2 \alpha_2 2 \, \r  ~ ,~
\nn  \hskip 1.30 cm  &(3.14 c,d)
\eea
\addtocounter{equation}{1}
where $\alpha_2 = 1,2 $ denotes the spin orientation of  2nd  particle
and $ f(1) = {g(1)}^{-1} = \epsilon_1 e^{2i\theta } $, 
$ f(2) = {g(2)}^{-1} = \epsilon_2 e^{2i\theta } $. It is rather
surprising to notice that, for $\epsilon_1 \neq \epsilon_2 $,
the phase factors in the expressions (3.14b,c) 
not only depend on the spin orientations of 1st and 3rd particle 
but also on that of the intermediate  2nd particle.
Therefore,  the operator ${\tilde P}_{13}$ no longer acts 
like identity on the 2nd internal space and generates a three-body
interaction when substituted to the CS Hamiltonian (\ref {c1}). 
It is worth noting that the form of ${\tilde P}_{12}$ in eqn.(\ref {c13})
is quite similar to the supersymmetric exchange operators [6,39]
associated with two species of particles, where 
$\epsilon_1 $ and $\epsilon_2$ play the role of corresponding
grading parameters. However, even in the supersymmetric case
the phase factors associated with 
a permutation operator like $P_{13}$ can depend only
on the gradings of spin components in the
1st and 3rd internal space, and are completely independent of the
spin orientation in the 2nd internal space. So the fact that the operator
$P_{13}$ given by eqn.(3.14) induces a three-body interaction
is rather unique to the present situation.

The above mentioned feature of `non-nearest neighbour' 
permutation operator $P_{13}$ becomes  even  more prominent
when,  by using the relations (\ref {c12}) and (\ref {c8}),
 one constructs the action of a  general 
element $ {\tilde P}_{kl}$ ($l-k > 1 $) on the related vector space
${\cal F}$ as
\bea
&&{\tilde P}_{kl} \,
 \v \alpha_1 \alpha_2 \cdots \alpha_k \cdots \alpha_l
\cdots \alpha_N \r ~~=~~~~~~~~~~~~~~~~~~~~~~~~~~~~~ \nn \\ &&  
 \quad \quad \quad
 \exp   \left \{  \,  i \, \phi_{\alpha_k \alpha_l} ~+~ i \,
\sum_{\tau =1}^M  \, n_\tau \,   \left (   \phi_{\tau \alpha_l} - 
 \phi_{\tau \alpha_k}    \right )  \,  \right \}   \,  
     \v \gamma_1
\gamma_2 \cdots \gamma_k \cdots \gamma_l \cdots \gamma_N \r   , ~~
\label {c15}
\eea
where $\gamma_k =\alpha_l,~\gamma_l =\alpha_k $ and $\gamma_j
= \alpha_j$ for $ j \neq k,l $;  we have  used the notation 
$e^{i \phi_{\sigma  \sigma }} = \epsilon_\sigma $,  
and assumed that the  particular spin orientation
$\alpha_p = \tau $ occurs $n_\tau $ number of times when the
index $p$ in $\alpha_p $ runs from $k+1$ to $l-1$. 
Thus,  the phase factor  associated with the
element ${\tilde P}_{kl}$ now depends on the spin configuration 
of $l-k+1$ number of particles:
$  \,\alpha_k, \, \alpha_{k+1},  \cdots ,  \, \alpha_l  $.  So, if we 
substitute this `anyon like' representation of permutation operators to  
 the spin CS Hamiltonian (\ref {c1}), it will lead to an
exactly solvable model with highly nonlocal interactions.
It is facinating to observe that such solvable models with nonlocal
interactions are quite similar to the quantum spin chains with open 
boundary conditions; since, from
eqn.(\ref {c15}) it may be seen that the  
spin dependent interaction between $1$st and $N$-th particle, i.e. 
$P_{1N}$, would be much more complicated in form
 than all other nearest neighbour spin interactions  like
 $P_{12},~P_{23}$ etc. 
This more complicated nature of operator $P_{1N}$ is probably
connected with the existence of some nonperiodic or twisted boundary
condition on the CS model. However, it should be noted
that the symmetry properties of these spin CS models are
completely different from that of 
the $BC_N$ type CS models [40-42], which  are well known
for their relevence in one dimensional physics 
with boundary. For example, the
particles in $BC_N$ type  CS model can  interact even 
with their `mirror images' and also with an impurity located
at the origin.  So these
interaction terms, depending in particular on the summation 
of coordinates like ($x_i + x_j$), break the 
translational invariance of the related system. On the other hand 
the CS Hamiltonians given by eqn.(\ref {c1}),  which 
depend only on the difference of particle coordinates, would
remain translational invariant even in the presence
of new types of spin-spin interactions.

\vspace{.75cm}

\noindent \section {Explicit solutions of different spin CS models}
\renewcommand{\theequation}{4.{\arabic{equation}}}
\setcounter{equation}{0}

\medskip
In the previous section we have seen that the
eigenfunctions of CS Hamiltonian 
(\ref {c1}), associated with an arbitrary representation of the 
permutation operators,  can be constructed by diagonalising the Dunkl operators
 and subsequently  using the general relation (\ref {c6}). 
Then we have also found some concrete examples of such exactly solvable
spin CS model, by inserting the anyon like representations (\ref {c12})
and (\ref {c15}) to the Hamiltonian (\ref {c1}). In the following
we like is to explicitly derive a few of the related 
eigenfunctions, by restricting to systems  which contain a 
small number of spin-${1\over 2}$  particles. 

For this purpose we first consider the simplest spin CS model,
which contain two spin-${1\over 2}$ particles moving on a circle.
The operator ${\tilde P}_{12}$, given by eqn.(\ref {c13}),
would represent the spin dependent
interaction between these two particles. Now,
by using eqn.(\ref {b11}), one may explicitly write down the
related Dunkl operators as
\beq 
d_1 ~=~  z_1 { \d \over \d z_1 } \, - \,  {\beta \over 2} \,  -  \,
\beta { z_1 \over z_1 -z_2 } ( K_{12} - 1 )~,~~
d_2 ~=~  z_2 { \d \over \d z_2 } \, +  \, {\beta \over 2} \,  +  \,
\beta { z_1 \over z_1 -z_2 } ( K_{12} - 1 )~.~~
\label {d1}
\eeq
So, according to our discussion in sec.2,  
 the trivial monomial with homogeneity zero (i.e. $  \xi 
= 1$\,) would be an eigenvector of the operators
$d_1, \, d_2 $ and will also correspond to 
the ground state of operator  
${\cal H}^* = d_1^2 + d_2^2 $. By using eqn.(\ref {d1}),
it is rather easy to see that the eigenvalues of the 
operators $d_1,~d_2 $ and ${\cal H}^*$ in this case will be
given by $- \beta / 2,$  $\beta / 2$ and $\beta^2 / 2$
respectively. Therefore, by applying the relation (\ref {c6}), 
the ground state of spin  CS Hamiltonian (\ref {c1}) 
associated with energy eigenvalue  $ { 2\pi^2 \over L^2 } \epsilon = 
{ \pi^2 \beta^2  \over  L^2 } $ 
can be found as
\beq
{\tilde \psi} ~~=~~\sin^\beta
 \left \{  { \pi \over L } (x_1 -x_2)   \right \}  ~
\left ( 1 - {\tilde P}_{12}  \right ) \rho( \alpha_1 , \alpha_2 ) ~, 
\label {d2}
\eeq
where 
$\rho (\alpha_1 , \alpha_2) $ is an arbitrary spin dependent function. 
 However,  for spin-${1\over 2}$  case
the function $\rho (\alpha_1 , \alpha_2 ) $ 
can be chosen in four different ways: $\v 11 \r $, 
$ \v  12 \r$,  $\v  21 \r$  and  $ \v 22 \r $. By inserting these
forms of $\rho$ to eqn.(\ref {d2}) and also using eqn.(\ref {c13}),
 we get three degenerate eigenfunctions like
$$
{\tilde \psi}_1  \, = \, ( 1 - \epsilon_1 ) \, \Gamma^\beta_{(2)} \v 11 \r~,~~
{\tilde \psi}_2 \, = \, ( 1 - \epsilon_2 ) \, \Gamma^\beta_{(2)} \v 22 \r~,~~
{\tilde \psi}_3 \, = \, \Gamma^\beta_{(2)} ~\left ( \v 12 \r - e^{i \theta }
 \v 21 \r \right )~,~ \eqno  (4.3a,b,c)
$$
\addtocounter{equation}{1} \noindent 
where $ \Gamma^\beta_{(2)} = \sin^\beta \left [ 
  { \pi \over L } (x_1 -x_2) \right ] $. Notice that the choice of 
$\rho $ as  $ \v 12 \r $ or  $\v 21 \r$ would  lead to
 the same wave function  $ {\tilde \psi}_3  $ up to a multiplicative constant.

Next, for constructing the first excited states of the above 2-body
problem, let us consider the monomials of homogeneity one. Evidently,
$z_2$ and $z_1$ are two such monomials
belonging to the partition $ \, (1+0) \, $  and $z_2$ is 
of higher order than $z_1$ according to the convention discussed in sec.2.
It is not difficult to check that for this simple case the
action of Dunkl operators (\ref {d1}) on  $z_2 , \, z_1$
will generate two simultaneously diagonalisable and triangular matrices,
whose eigenvectors (i.e. $z_2'$ and $z_1$) would
satisfy the relations
\beq
d_1   z'_2 ~=~  - \, { \beta \over 2 } \,  z'_2 ~,~~
d_2 z'_2 ~=~  \left ( 1 + { \beta \over 2 } \right ) z'_2 ~,~~
d_1 z_1 ~=~  \left ( 1 + { \beta \over 2 } \right ) z_1 ~,~~
d_2 z_1 ~=~  - \, { \beta \over 2 } \,  z_1 ~,
\label {d4}
\eeq
where $z'_2 = ( 1+\beta ) z_2 + \beta z_1 $. Consequently,
$ \epsilon = \beta^2 /2 + \beta + 1$ will be the eigenvalue 
of the operator ${\cal H}^*$ for both of the eigenstates
$z'_2$ and $z_1$.
So, by using (\ref {c6}), 
we find that the first excited states of spin CS 
Hamiltonian (\ref {c1}) are given by the expressions
$$
{\tilde \psi}' ~=~
 \Gamma^\beta_{(2)}  \left ( 1 - K_{12} {\tilde P}_{12} \right ) 
\rho (\alpha_1 , \alpha_2 ) z_1 ~,
~~~ {\tilde \psi}'' ~=~
 \Gamma^\beta_{(2)}  \left ( 1 - K_{12} {\tilde P}_{12} \right ) 
\rho (\alpha_1 , \alpha_2 )  z_2' ~,
\eqno (4.5a,b)
$$
\addtocounter{equation}{1}
where, as before,  $\rho (\alpha_1 , \alpha_2 )  $ 
is an arbitrary spin dependent function. 
If we insert four possible choice of $\rho $ 
to eqn.(4.5a), that would lead to four degenerate wave functions like
\bea
&& ~~~~{\tilde \psi}_1' ~=~
 \Gamma^\beta_{(2)}  ~( z_1 - \epsilon_1 z_2 ) \v 11 \r ~,~~
{\tilde \psi}_2' ~=~
 \Gamma^\beta_{(2)}  ~( z_1 - \epsilon_2 z_2 ) \v 22 \r ~,~~\nn \\
&& {\tilde \psi}_3' \, = \, 
 \Gamma^\beta_{(2)}  \, ( z_1 - z_2 ) \left [\, \v 12 \r  + e^{i \theta }
\v 21 \r \, \right ]\, ,~
{\tilde \psi}_4' \, = \, 
 \Gamma^\beta_{(2)}  \, ( z_1 + z_2 ) \left [ \,  \v 12 \r  - e^{i \theta }
\v  21  \r \, \right ] , ~~~~~~ 
\label {d6}
\eea
which naturally share the same energy  eigenvalue  
$ { 2\pi^2 \over L^2 } \epsilon = 
{ \pi^2  \over  L^2 }  ( \beta^2 + 2 \beta + 2 ) $.
However, if we substitute the four different
 forms of $\rho (\alpha_1 , \alpha_2 ) $ to
eqn.(4.5b), that will only reproduce the above four wave functions.

Now we like to analyse the effect of different
permutation operators on the above constructed ground state
and first excited state wave functions, 
by tuning the discrete parameters $\epsilon_1,~\epsilon_2 $ as well as
the continuous parameter $\theta $. For this purpose,
 first we insert $\epsilon_1 = \epsilon_2 = 1 $
to the eqns.(4.3), (\ref {d6}) and curiously notice that some of
the wave functions appearing in  eqn.(4.3)  would become trivial
for these values of discrete parameters. So we write down
  the explicit form of the remaining nontrivial wave functions as
\bea
&& \psi_{\bf 0}^{(1)}  = \, \Gamma^\beta_{(2)} \,
\left (  \v 12 \r - e^{i \theta }
 \v 21 \r  \right ) \, ,~ \psi_{\bf 1}^{(1)} = \,
 \Gamma^\beta_{(2)}  \, ( z_1 - z_2 ) \v 11 \r \, ,~\psi_{\bf 1}^{(2)} = \,
 \Gamma^\beta_{(2)}  \, ( z_1 - z_2 ) \v 22 \r  \, ,\nn \\
&& \psi_{\bf 1}^{(3)}  =  
 \Gamma^\beta_{(2)}  \, ( z_1 - z_2 ) \left ( \, \v 12 \r  + e^{i \theta }
\v 21 \r \, \right ) \,  ,~ \psi_{\bf 1}^{(4)}   = 
 \Gamma^\beta_{(2)}  \, ( z_1 + z_2 ) \left ( \,  \v 12 \r  - e^{i \theta }
\v  21  \r \, \right ) \,  ,~
\label {d7}
\eea
which shows the existence of a nondegenerate ground state 
$ \psi_{\bf 0}^{(1)} $ with energy $\pi^2 \beta^2 / L^2 $
and a four-fold degenerate first excited state
$ \psi_{\bf 1}^{(i)} $ with energy
 $\pi^2 ( \beta^2 + 2\beta + 2 ) / L^2 $.  It was mentioned earlier
 that for the values $\epsilon_1 = \epsilon_2 = 1 $
 and $\theta =0 $, the permutation operator
(\ref {c13}) coincides with its standard counterpart  and 
yields the $su(2)$ invariant CS model (\ref {a1}). Consequently,
the ground state and first excited states of this $su(2)$ invariant
model can be easily reproduced
 by simply putting $\theta = 0$ in the expression (\ref {d7}).

Next we take the values of our discrete paramaters as
 $\epsilon_1 = - \epsilon_2 =  1 $, 
which is related to a new variant of spin CS model. Again, by
 inserting these values to
(4.3) and (\ref {d6}), one may find out the corresponding nontrivial
 wavefunctions as
\bea
&& \psi_{\bf 0}^{(1)} \, = \, 2 \Gamma^\beta_{(2)} \, \v 22 \r  \, ,~~
 \psi_{\bf 0}^{(2)} \, = \,  \Gamma^\beta_{(2)} \,
 ( \, \v 12 \r - e^{i \theta } \v 21 \r \, ) \, ,~~ \nn \\
&& 
\psi_{\bf 1}^{(1)} \, = \,
 \Gamma^\beta_{(2)}  \, ( z_1 - z_2 ) \v 11 \r  \,  ,
~~ \psi_{\bf 1}^{(2)} \, = \,
 \Gamma^\beta_{(2)}  \, ( z_1 + z_2 ) \v 22 \r  \, ,~~
 \psi_{\bf 1}^{(3)} \,  ,~~ \psi_{\bf 1}^{(4)} \,  ,
\label {d8}
\eea
where the forms of
 $ \psi_{\bf 1}^{(3)}  $  and $ \psi_{\bf 1}^{(4)} $ are
 identical to their previous forms which appeared in  
eqn.(\ref {d7}). So for these values of discrete parameters 
and the related nonstandard spin CS model,
one gets a doubly degenerate ground state along with a four-fold degenerate
first excited state. Similarly, for the values of discrete parameters as
$\epsilon_1 = - \epsilon_2 = - 1 $, it is easy to
  see that the  associated  wave functions would be given by
\bea
&& \psi_{\bf 0}^{(1)} \, = \, 2 \Gamma^\beta_{(2)} \, \v 11 \r  \, ,~~
 \psi_{\bf 0}^{(2)} \, = \,  \Gamma^\beta_{(2)} \,
 ( \, \v 12 \r - e^{i \theta } \v 21 \r \, ) \,  ,~~ \nn \\
&& 
\psi_{\bf 1}^{(1)} \, = \,
 \Gamma^\beta_{(2)}  \, ( z_1 + z_2 ) \v 11 \r  \, ,~~
 \psi_{\bf 1}^{(2)} \, = \,
 \Gamma^\beta_{(2)}  \, ( z_1 - z_2 ) \v 22 \r  \, ,~~
 \psi_{\bf 1}^{(3)}  ~,~~ \psi_{\bf 1}^{(4)}  \, ,
\label {d9}
\eea
which again shows
 a doubly degenerate ground state and a four-fold degenerate
first excited state.  However, it is worth noting 
that the forms of present eigenfunctions
 $ \psi_{\bf 0}^{(1)} ,~ \psi_{\bf 1}^{(1)}$ and $ ~ \psi_{\bf 1}^{(2)}$
are quite different from their respective
forms in eqn.(\ref {d8}) associated with 
  $\epsilon_1 = - \epsilon_2 =  1 $ sector.
Finally, one may also find out 
the wave functions corresponding to  the  case
 $\epsilon_1 =  \epsilon_2 = - 1 $ as
\bea
&& \psi_{\bf 0}^{(1)} \, = \, 2 \Gamma^\beta_{(2)} \, \v 11 \r ~,~~
\psi_{\bf 0}^{(2)} \, = \, 2 \Gamma^\beta_{(2)} \, \v 22 \r ~,~~
 \psi_{\bf 0}^{(3)} \, = \,  \Gamma^\beta_{(2)} \,
 ( \v 12 \r - e^{i \theta } \v 21 \r )~,~~ \nn \\
&& 
\psi_{\bf 1}^{(1)} \, = \,
 \Gamma^\beta_{(2)}  \, ( z_1 + z_2 ) \v 11 \r  \, ,
~~ \psi_{\bf 1}^{(2)} \, = \,
 \Gamma^\beta_{(2)}  \, ( z_1 + z_2 ) \v 22 \r  \, ,~~
 \psi_{\bf 1}^{(3)}  \, ,~~ \psi_{\bf 1}^{(4)}  \, .~~
\label {d10}
\eea
So, in this sector, one interestingly gets 
 a triply degenerate ground state along with a four-fold degenerate
first excited state.

Thus from the nature of
above construction it is clear that, the
ground state  energy and the first excited state energy
of 2-particle spin CS model do not depend on the
choice of parameters $\epsilon_1 ,~ \epsilon_2 $ and $\theta $
in the related Hamiltonian. However, the values of 
discrete parameters $\epsilon_1 $ and $\epsilon_2$
can affect the degeneracy of the
ground state in a very  significant way. While the 
standard choice $\epsilon_1 = \epsilon_2 = 1 $ yields a 
nondegenerate ground state, other possible choice of these 
two discrete parameters  would give us a doubly or triply degenerate
ground state. On the other hand,
this degeneracy factor does not change at all with the variation of  
continuous parameter $\theta  $. So, only the explicit
form of these ground state wave functions, and not their degeneracy
factor, would depend on the value of $\theta $. Furthermore
it turns out that, in contrast  to the  case of ground  state, 
the first excited 
state always remain four-fold degenerate for any possible choice of 
the parameters $\epsilon_1 ,~ \epsilon_2 $ and $\theta $.

It is easy to similarly  derive
the wave functions and their degeneracy factors related to
the higher excitations  of two spin-${1\over 2}$ particles.
However, in the following, we like to focus our attention to the
CS model containing three
spin-${1\over 2}$ particles and  explore whether such a system
exhibits any new interesting feature. In this case, we have to
simultaneously diagonalise three Dunkl operators $d_1$, $d_2$ and
$d_3$, which can be explicitly written by using the relation
(\ref {b11}). It is easy to check that, 
 the trivial monomial with homogeneity zero ($\xi =1 $)
would be the simplest eigenstate of $d_1$, $d_2$ and $d_3$ with 
eigenvalues  $ - \beta , ~0,$ and $\beta $ respectively. Moreover,
 due to the relation (\ref {b10}), $\epsilon = 2 \beta^2 $ will be the 
eigenvalue for the operator ${\cal H}^*$ corresponding to this eigenstate. 
  So, by using  eqn.(\ref {c6}), one can find out
the ground state of CS model (\ref {c1}) as
\beq
{\tilde \psi} ~~=~~\Gamma^\beta_{(3)} \, {\tilde \Lambda }_{3} \,
 \rho (\alpha_1 , \alpha_2 , \alpha_3 )
\label {d11}
\eeq
where 
$ \Gamma^\beta_{(3)} ~=~ \sin^\beta \left [ {\pi \over L}  (x_1 -x_2 )
\right ] \sin^\beta \left [ {\pi \over L}  (x_2 -x_3) \right ] $,
  and the `generalised'
 antisymmetric projection operator ${\tilde \Lambda }_{3}$ is 
given by the expression (3.5b) which  at present  contains the
representations of permutation operators like
(\ref {c13}) and (3.14). Furthermore, by taking
the arbitrary spin-dependent function  $\rho (\alpha_1 , \alpha_2 ,
\alpha_3 ) $ in the above equation
in  eight possible ways: $ \v 111 \r $, $\v 112 \r $,
$\v 121 \r $, $\v 122 \r $ 
$\v  211 \r $, $\v  212 \r $,  $\v  221 \r $  and $\v  222 \r$,
we obtain four distinct eigenfunctions given by
\bea
&& {\tilde \psi}_1 \, = \, ( 1- \epsilon_1 )  \, \Gamma^\beta_{(3)}
\v 111 \r ,~ \, 
 {\tilde \psi}_2 \, = \, ( 1- \epsilon_1 ) \, \Gamma^\beta_{(3)}
\, \left ( \,  \v 112 \r - e^{i\theta } \v 121 \r  +
 e^{2i\theta } \v  211  \r \,  \right ) ,~~~~~ \nn \\
&& {\tilde \psi}_3 \,  = \,  ( 1- \epsilon_2 ) \, \Gamma^\beta_{(3)}  \,
\v  222 \r ,~ \, 
{\tilde \psi}_4 \,  = \,  ( 1- \epsilon_2 ) \, \Gamma^\beta_{(3)}
\left ( \,  \v  122 \r  - e^{i \theta } \v  212 \r  + 
e^{2i\theta } \v  221 \r  \,  \right )  . ~~~~~
\label {d12}
\eea
It is evident that
 ${ 2\pi^2 \over L^2 } \epsilon = { 4 \pi^2 \beta^2 \over L^2 }$
would be the energy eigenvalue for all of these degenerate states. 

Next we consider the monomials $z_3, ~z_2$ and $z_1$, which  correspond
to  the partition $~(1+0+0) \,$ of  homogeneity one  sector. In this case 
one can again simultaneously diagonalise the triangular
matrix representations which are generated by the action of Dunkl 
operators on these three monomials  
and obtain the related eigenvectors
as
\beq 
z_3' ~=~ \beta \left ( z_1 + z_2 \right )  + ( 1+ \beta ) z_3 ~,~~
z_2' ~=~ \beta z_1 + ( 1+2\beta ) z_2 ~,~~
z_1' ~=~ z_1~.
\label {d13}
\eeq
So,  with the help of eqn.(\ref {c6}), we find  that
the first excited states of spin CS model (\ref {c1}) 
would be given by the expressions: $~{\tilde \psi }' = \Gamma_{(3)}^\beta 
{\tilde \Lambda }_{(3)} \left ( \, \rho z_1' \, \right ) $,
 $ {\tilde \psi }'' = \Gamma_{(3)}^\beta 
{\tilde \Lambda }_{(3)} \left ( \, \rho z_2' \, \right ) $,
and $ {\tilde \psi }''' = \Gamma_{(3)}^\beta 
{\tilde \Lambda }_{(3)} \left ( \, \rho z_3' \, \right ) $.
By inserting the previously mentioned eight possible forms
of the arbitrary function  $ \rho (\alpha_1 , \alpha_2 , \alpha_3 ) $
 to ${\tilde \psi }' $, we obtain a set of six distinct and degenerate 
eigenfunctions like
\bea
&& { \tilde \psi }'_1 ~=~ ( 1- \epsilon_1 ) \, \Gamma_{(3)}^\beta 
\left ( z_1 + z_2 + z_3 \right ) \v 111 \r ~,~~
 { \tilde \psi }'_2 ~=~ ( 1- \epsilon_2 ) \, \Gamma_{(3)}^\beta 
\left ( z_1 + z_2 + z_3 \right ) \v  222 \r  \, ,~ 
\nn \\ 
&&{ \tilde \psi }'_3 ~=~ ( 1- \epsilon_1 ) \, \Gamma_{(3)}^\beta 
 \left \{ ~  z_1 \v  211  \r  -  e^{-i \theta } z_2 \v  121 \r 
+ e^{-2i \theta  } z_3 \v  112 \r ~ \right  \}  \,  ,~~~ ~~
\nn  \\
&&{ \tilde \psi }'_4 ~=~ ( 1- \epsilon_2 ) \, \Gamma_{(3)}^\beta 
\left \{  ~  z_1 \v  122  \r  - e^{i \theta } z_2 \v  212  \r
+ e^{2i \theta } z_3 \v 221 \r ~ \right  \}  \, ,~
\nn  \\
&&{ \tilde \psi }'_5 ~ = ~ \Gamma_{(3)}^\beta 
\left \{  ~  (z_1 - \epsilon_1 z_2) \v  112  \r
-  e^{i \theta } (z_1 - \epsilon_1 z_3) \v 121  \r
+  e^{2i \theta } (z_2 - \epsilon_1 z_3) \v 211 \r ~ \right \}  \, ,~
\nn \\
&& { \tilde \psi }'_6 ~ = ~ \Gamma_{(3)}^\beta 
\left  \{ ~ (z_1 - \epsilon_2 z_2) \v 221  \r
-  e^{- i\theta } (z_1 - \epsilon_2 z_3) \v 212  \r 
+  e^{-2i \theta } (z_2 - \epsilon_2 z_3) \v 122  \r ~ \right \}  \,
. \nn \\
&&  \hskip  11.25  cm  \label {d14}
\eea
 It may be noted that if one substitutes
the eight possible forms of arbitrary function $\rho $ to
 ${\tilde \psi }''$ or  $ {\tilde \psi }''' $, that  will only
reproduce the above set of six wave functions.
Moreover, by using eqns.(\ref {b10}) and (\ref {b11}), 
it is easy to check that 
$ \, \epsilon = 3\beta^2 + 2\beta + 1 \, $ would be 
the eigenvalue of operator ${\cal H}^*$ for all  $z_i'$ in 
eqn.(\ref {d13}). Consequently, the degenerate wave functions 
appearing in eqn.(\ref {d14}) 
will share the same energy eigenvalue $ 2 \pi^2 
 (3\beta^2 + 2\beta + 1) \ L^2 $.  

Now, similar to the  case of  two particles, let us
  analyse again the effect of different
permutation operators on the above constructed ground state
and first excited state wave functions  associated with 
three spin-${1\over 2}$ particles. In this context it 
is interesting to observe that,
all four wave-functions in eqn.(\ref {d12}) would  vanish identically
for the choice of two discrete parameters as   
$\epsilon_1 = \epsilon_2 = 1 $.
This observation is also consistent with the simple fact that
the usual antisymmetrisation of more than two spin-${1\over 2}$ particles 
always yields the trivial result.  So,
to  obtain the related ground state,  it is necessary to consider the 
monomials of  homogeneity  one instead of homogeneity zero. Therefore,
we insert the values $\epsilon_1 = \epsilon_2 = 1 $ to eqn.(\ref {d14}) 
and find that there exist two nontrivial wave functions
which may be explicitly written as
\bea 
&& \psi_{\bf 0}^{(1)}  \,  =  \,  \Gamma_{(3)}^\beta 
\left \{ ~(z_1 - z_2) \v  112  \r
-  e^{i \theta } (z_1 - z_3) \v 121  \r
+  e^{2i \theta } (z_2 - z_3) \v 211 \r ~ \right \}  \, ,~~
\nn \\
&& \psi_{\bf 0}^{(2)}  \,  =  \,  \Gamma_{(3)}^\beta 
\left  \{ ~ (z_1 - z_2) \v 221  \r
-  e^{- i\theta } (z_1 -  z_3) \v 212  \r 
+  e^{-2i \theta } (z_2 - z_3) \v 122  \r ~ \right \}   .~~~ 
\label {d15}
\eea
Thus, for these values of discrete parameters, one gets a 
doubly degenerate ground state with energy eigenvalue 
$2 \pi^2 (3\beta^2 + 2\beta + 1) / L^2$. It may be 
noticed that the above 
equation will also reproduce the ground state of usual 
$su(2)$ invariant spin CS model (\ref {a1}), after the
substitution $\theta = 0$.

Next we take the values of discrete parameters as 
$\epsilon_1 = - \epsilon_2 = 1 $,  which would lead to 
a nonstandard type of spin CS model. By putting these values of 
$\epsilon_1 $ and $ \epsilon_2 $ to eqns.(\ref {d12})
 and (\ref {d14}) respectively, it is straightforward to 
find that for such nonstandard  spin CS model
 there exist a doubly degenerate ground state 
with energy $4 \pi^2 \beta^2 /L^2$ :
\beq
\psi_{\bf 0}^{(1)} \,  = \,  2 \, \Gamma^\beta_{(3)}  \,
\v  222 \r ,
~~~ \psi_{\bf 0}^{(2)} \,  = \,  2  \, \Gamma^\beta_{(3)}
  \left ( \,  \v  122 \r  - e^{i \theta } \v  212 \r  + 
       e^{2i\theta } \v  221 \r  \,  \right )  ~ ,
\label {d16}
\eeq  
and a four-fold degenerate first excited state with 
energy $2 \pi^2 (3\beta^2 + 2\beta + 1) / L^2$ :
\bea 
&&
 \psi_{\bf 1}^{(1)}  = 2  \Gamma_{(3)}^\beta 
\left ( z_1 + z_2 + z_3 \right ) \v  222 \r  \,  ,  ~
\psi_{\bf 1}^{(2)}  =   2  \Gamma_{(3)}^\beta 
\left \{   z_1 \v  122  \r  - e^{i \theta } z_2 \v  212  \r
+ e^{2i \theta } z_3 \v 221 \r  \right  \}  \, , 
\nn  \\
&&
\psi_{\bf 1}^{(3)} \,  =  \, \Gamma_{(3)}^\beta 
\left \{  ~  (z_1 -  z_2) \v  112  \r
-  e^{i \theta } (z_1 -  z_3) \v 121  \r
+  e^{2i \theta } (z_2 - z_3) \v 211 \r ~ \right \}  \, ,
\nn \\
&&
\psi_{\bf 1}^{(4)} \,  =  \,  \Gamma_{(3)}^\beta 
\left  \{ ~ (z_1 +  z_2) \v 221  \r
-  e^{- i\theta } (z_1  + z_3) \v 212  \r 
+  e^{-2i \theta } (z_2  +  z_3) \v 122  \r ~ \right \}  \, .
\label {d17}
\eea
By using eqns.(\ref {d12}) and (\ref {d14}), one can similarly obtain 
the ground states and first excited states of spin 
CS models associated with other values of the two discrete parameters.
In particular, it is easy to see that
 the sector $\epsilon_1 = - \epsilon_2 = - 1 $ would again lead to 
a doubly degenerate ground state and a four-fold degerate first excited state. 
On the other hand, the remaining sector $\epsilon_1 =  \epsilon_2 = - 1 $
would give us a four-fold degenerate ground state and a six-fold degenerate 
first excited state. 

It is rather interesting to notice that the energy of 
 ground states appearing in eqn.(\ref {d16}) is actually lower
 than that of the previous ground states (\ref {d15}). 
Consequently, the  choice of two discrete parameters as
 $\epsilon_1 = - \epsilon_2 = \pm 1  $ and 
 $\epsilon_1 =  \epsilon_2 = - 1  $,  
would provide  us nonstandard variants  of spin CS model whose
ground state energy is lower than that of the 
usual spin CS model (\ref {a1}) associated with 
 the $\epsilon_1 = \epsilon_2 = 1 $ sector. Furthermore,
the ground state energy of usual spin CS model turns out to be exactly
same with the first excited state energy of all other nonstandard
variants of this model.
Thus we curiously find that
for the CS model containing three spin-${1\over 2}$ particles,
  it is not only possible to change the degeneracy factor 
of the ground state, but also its energy level, by tuning two
discrete parameters which appear
 in the anyon like representations of permutation operators.
\vspace{.75cm}
\noindent \section { Concluding Remarks }

Here we carefully analyse the 
method of constructing solutions of  spin dependent
Calogero-Sutherland (CS) model and observe that 
the algebra of the permutation operators,
rather than any of  their particular representation,
plays an important role   in this context.
Moreover we consider a $A_{N-1}$ type of spin CS model,  containing 
an arbitrary representation of  permutation operators 
BB
on the combined internal space (${\cal F}$) of all particles, 
 and find that such a model can be solved almost in the same way as
 its standard $su(M)$ invariant counterpart by  introducing a 
`generalised' antisymmetric projection operator. 

Next, with the aim of
constructing new variants of spin CS model, 
we search for  some explicit representations of permutation
operators on the space ${\cal F}$. Here we interestingly
notice that a class of known representations of the Hecke algebra, 
characterised by a deformation parameter $q$, 
reduces to such representations at the limit $ q\rightarrow 1$. 
These representations  of permutation operator ${\cal P}_{kl}$ ($k<l$)
not only interchange the spins of $k$-th and $l$-th  particles,
but also pick up nontrivial phase factors depending on the
spin  configuration of all particles indexed by
$k, ~k+1, \cdots ,~l$. 
Moreover  these  `anyon like' representations are found to be 
dependent on ${ M (M-1) \over 2}$ number of  continuously variable
 antisymmetric parameters $\phi_{\gamma \sigma}$, as well as $M$ number of 
discrete parameters $\epsilon_\sigma $ which 
can be freely chosen to be $ 1$ or $-1$. At the special case
$\phi_{\gamma \sigma} = 0 $ and
$\epsilon_\sigma = 1 $ for all values of $\gamma ,~ \sigma $,
they   coincide with the usual  representation 
of permutation operator,  which only interchanges the spins of 
two particles and  leads to 
 the standard $su(M)$ invariant CS model. However,  other
possible values  of the  parameters 
$\epsilon_\sigma $ and $\phi_{\gamma \sigma}  $ would  generate
  novel  variants of spin  CS  Hamiltonian,   containing 
highly nonlocal type  of spin dependent 
interactions,  which  violate the  $su(M)$ invariance.

Subsequently,  we explicitly derive
 a few low-lying energy states  of the above mentioned spin CS models,
by restricting to systems which 
contain a small number of  spin-${1\over 2}$  particles. 
For the case of two spin-${1\over 2}$ particles,  we find that there 
exists a non-degenerate ground state associated with the choice of discrete
parameters  $\epsilon_1 = \epsilon_2 = 1$.  However, other possible values of
the  parameters $\epsilon_1$ and $\epsilon_2$  curiously
yield 2-fold or 3-fold degenerate ground state with the same energy 
level. Thus it turns out that,  the choice 
of discrete parameters in the representations of permutation operators
can affect  the degeneracy of the 
related ground states in a significant way.  On the other hand,
this degeneracy factor is found to be insensitive to the value
of the continuous parameter $\phi_{12} $.
More interesting things happen if one  considers  a system
with three spin-${1\over 2}$ particles. In this case 
we find that, both  the degeneracy factor of the ground state as well as its
 energy level crucially depend on the choice of two discrete
parameters $\epsilon_1 $ and $\epsilon_2 $. In fact, the 
 ground state energy associated with the sector
$\epsilon_1 = \epsilon_2 = 1$ exactly coincides  with
the  energy of first  excited states associated with 
$\epsilon_1 = \epsilon_2 =- 1$ and $\epsilon_1 =- \epsilon_2 = \pm 1$
sectors.

The approach presented here for constructing
novel types of spin CS models might have some
further implications in several directions. 
As it is well known, the $su(M)$ invariant  Haldane-Shastry model
is related to a `frozen' limit of the spin CS model [7].
So it should be encouraging to explore whether 
the Hamiltonian of this exactly solvable Haldane-Shastry
model can also  be  modified
through our anyon like representations of permutation operators. 
Moreover,  it might be fruitful to investigate about various dynamical 
correlation functions and thermodynamic quantities of such new models
in connection with the fractional statistics.
Another relevant  problem   is to   establish the integrability
of different spin  CS models which are discussed in this  article
and find out the algebra of corresponding  conserved quantities.
By investigating along this line 
we have observed very recently that [43]  a multi-parameter  
 dependent extension  of $Y(gl_N)$ Yangian [44,45], 
as well as  its `nonstandard' variants,  curiously
play the role  of symmetry algebra for these CS models. 
It may be hoped that the representation 
theory of such extended Yangian algebra would give us 
some valuable insight about the degeneracy factors of the related 
quantum states.

\medskip
\noindent {\bf Acknowledgments }

It is a pleasure to thank Profs. Anjan Kundu, Sunil Mukhi 
 and B.S. Shastry for many illuminating discussions. The author
is also grateful to the Referee for his constructive suggestions 
which helped to improve this paper considerably.
\newpage 
\leftline {\large \bf References } 
\medskip 
\begin{enumerate}

\item  F. Calogero, J. Math. Phys. 10 (1969) 2191.

\item  B. Sutherland, Phys. Rev. A 5 (1972) 1372.

\item  Z.N.C. Ha and F.D.M. Haldane, Phys. Rev. B 46 (1992) 9359.

\item  J.A. Minahan and A. P. Polychronakos, Phys. Lett. B 302 (1993) 265.

\item  K. Hikami and M. Wadati, J. Phys. Soc. Jpn. 62 (1993) 469.

\item  B. Sutherland and B.S. Shastry, Phys. Rev. Lett. 71 (1993) 5.

\item  D. Bernard, M. Gaudin, F.D.M. Haldane and V. Pasquier,
       J. Phys. A 26 (1993) 5219.

\item  V. Pasquier, {\it A lecture on the Calogero-Sutherland models},
       Saclay Preprint (1994) SPhT/94-060, hepth/9405104.

\item  F.D.M. Haldane, in Proc. 16th Taniguchi Symp., Kashikijima,
       Japan, (1993) eds. A. Okiji and N. Kawakami (Springer, Berlin,
       1994).

\item  Z.N.C. Ha, Phys. Rev. Lett. 73 (1994) 1574;  Nucl. Phys. B 435 [FS]
       (1995) 604.

\item  M.V.N. Murthy and R. Shankar, Phys. Rev. Lett. 73 (1994) 3331.

\item  F. Lesage, V. Pasquier and D. Serban, Nucl. Phys. B 435 [FS] (1995)
       585. 

\item  D. Sen, {\it A multispecies Calogero-Sutherland  model and mutual 
       exclusion statistics  }  (1995) Cond-mat/9512014.

\item  N. Kawakami, Phys. Rev. Lett. 71 (1993) 275.

\item  H. Azuma and S. Iso, Phys Lett. B 331 (1994) 107.

\item  M. Stone and M. Fisher, Int. J. Mod. Phys. B 8 (1994) 2539.

\item  B.D. Simons, P.A. Lee and B.L. Altshuler,  Nucl. Phys. B 409
(1993) 487.

\item  O. Narayan and B.S. Shastry, Phys. Rev. Lett. 91 (1993) 2106.

\item  F.D.M. Haldane and M. Zirnbauer, Phys. Rev. Lett. 71 (1994) 4055.

\item  A. Jevicki, Nucl. Phys. B 376 (1992) 75.

\item  H. Awata, Y. Matsuo, S. Odake and J. Shiraishi, Phys. Lett. B 347
(1995) 49.

\item  K. Hikami and M. Wadati, Phys. Rev. Lett. 73 (1994) 1191.

\item  H. Ujino and M. Wadati, J. Phys. Soc. Jap. 63 (1994) 3585;
J. Phys. Soc. Jpn. 64 (1995) 39.

\item  H. Awata, Y. Matsuo, S. Odake and J. Shiraishi, Nucl. Phys. B 449
(1995) 347.

\item  C.F. Dunkl, Trans. Am. Math. Soc. 311 (1989) 167.

\item  A.P. Polychronakos, Phys. Rev. Lett. 69 (1992) 703.

\item  L. Brink, T.H. Hansson and M. Vasiliev, Phys. Lett. B 286 (1992) 109.

\item  M. Wadati, T. Deguchi, and Y. Akutsu, Phys. Rep. 180 (1989) 247.

\item  L.D. Faddeev, Int. J. Mod. Phys. A 10 (1995) 1845.

\item  B. Basu-Mallick and A. Kundu, J. Phys. A 25 (1992) 4147.

\item  V.G. Drinfeld, {\it Quantum Groups }, in ICM proc. (Berkeley, 1987)
p. 798.  

\item L.D. Faddeev, N. Yu. Reshetikhin, and L.A. Takhtajan, in  
Yang-Baxter Equation in Integrable systems, Advanced series in Math. Phys. 
Vol. 10, edited by M. Jimbo ( World Scientific, Singapore, 1990) p.299.  

\item  V. Chari and A. Pressley,  A  Guide to Quantum Groups (Cambridge 
Univ. Press, Cambridge, 1994).

\item  M.L. Ge, C.P. Sun and K. Xue, Int. J. Mod. Phys. A 7 (1992) 6609.

\item  R. Chakrabarti and R. Jagannathan, Z. Phys. C 66 (1995) 523.

\item  S. Majid and M.J. Rodriguez-Plaza, J. Math. Phys. 36 (1995) 7081.

\item  A. Schirrmacher, Z. Phys. C 50 (1991) 321.

\item  J.H.H. Perk and C.L. Schultz, Phys. Lett. A 84 (1981) 407.

\item  C. Ahn and W.M. Koo, Phys. Lett. B 365 (1996) 105.
  
\item  M.A. Olshanetsky and A.M. Perelomov, Phys. Rep. 94 (1983) 313.

\item  A. Kapustin and S. Skorik, Phys. Lett. A 196 (1994) 47.

\item  T. Yamamoto, N. Kawakami and S.K. Yang, J. Phys A 29 (1996) 317.

\item  B. Basu-Mallick and Anjan Kundu, Under Preparation.

\item  B. Basu-Mallick and P. Ramadevi, Phys. Lett. A 211 (1996) 339.

\item  B. Basu-Mallick, P. Ramadevi and R. Jagannathan, {\it  Multiparametric
and coloured extensions of the quantum group $GL_q(N)$ and the Yangian 
algebra $Y(gl_N)$ through a symmetry transformation of the Yang-Baxter 
equation},(1995) q-alg/9511028.

\end{enumerate} 
\end{document}